\documentclass[11pt]{article}
\usepackage[margin=1in]{geometry}
\usepackage{amsmath,amssymb}
\usepackage{booktabs,array,longtable}
\usepackage{fancyvrb}
\usepackage{hyperref}
\usepackage{xcolor}
\usepackage{orcidlink}
\hypersetup{colorlinks=true,linkcolor=blue!50!black,citecolor=blue!50!black,urlcolor=blue!50!black}
\allowdisplaybreaks

\newcommand{\Z}{\mathbb{Z}}

\newcommand{\one}{\mathbf{1}}
\newcommand{\MM}{M_{\langle 3,3,3\rangle}}

\title{An Exact 56-Addition, Rank-23 Scheme for General $3\times3$ Matrix Multiplication}
\author{\textbf{Yinqi Sun}~\orcidlink{0009-0000-3783-5472}\\
        Department of Computer Science\\
        Purdue University\\
        \texttt{sun1226@purdue.edu}}
\begin{document}
\maketitle

\begin{abstract}
We present a rank-$23$ algorithm for general $3\times3$ matrix multiplication that uses $56$ additions/subtractions and $23$ multiplications, for a total of $79$ scalar operations in the standard bilinear straight-line model. This improves the recent sequence of $60$-, $59$-, and $58$-addition rank-$23$ schemes. The algorithm works over arbitrary associative, possibly noncommutative, coefficient rings. Its tensor coefficients are ternary, meaning that every coefficient lies in $\{-1,0,1\}$. Correctness is certified by the $729$ Brent equations over $\Z$, and the verifier also expands the straight-line program and performs additional finite-field
and noncommutative implementation tests.
\end{abstract}

\noindent\textbf{Keywords.} Fast matrix multiplication; bilinear algorithms; tensor rank; additive complexity.

\section{Introduction}

Strassen's algorithm showed that matrix multiplication can be accelerated by reducing the number of bilinear products rather than evaluating the defining sums directly \cite{Strassen1969}.  For $3\times3$ matrix multiplication over noncommutative rings, Laderman found a rank-$23$ algorithm in 1976 \cite{Laderman1976}.  That rank remains the smallest known rank for general $3\times3$ multiplication, while recent work has reduced the number of additions required to realize rank-$23$ schemes \cite{Smirnov2013,KarstadtSchwartz2020,BeniaminiEtAl2020,MartenssonWagner2025,Stapleton2025,MartenssonWagnerStapleton2025,Perminov2025}.

This paper records an explicit $56$-addition realization of such a rank-$23$ scheme.  The contribution is the straight-line program in Section~\ref{sec:algorithm}: it can be implemented directly from the printed formulas, without consulting an external certificate.  Appendix~\ref{app:expanded} gives an expanded bilinear view, Appendix~\ref{app:fileformat} gives the same factors in compact ternary file format, and Appendix~\ref{app:reproducibility} includes a github repository that contains our implementation of the algorithm and verification code for independent checking.

However, we do not claim that $56$ additions is globally optimal among all rank-$23$ schemes, nor that the displayed output-side straight-line program is globally optimal among all linear programs with cancellation.
\section{Model and notation}
\label{sec:model}

We use the row-major indexing convention
\[
C=AB=\begin{pmatrix}C_0&C_1&C_2\\ C_3&C_4&C_5\\ C_6&C_7&C_8\end{pmatrix}
=\begin{pmatrix}A_0&A_1&A_2\\ A_3&A_4&A_5\\ A_6&A_7&A_8\end{pmatrix}
 \begin{pmatrix}B_0&B_1&B_2\\ B_3&B_4&B_5\\ B_6&B_7&B_8\end{pmatrix}.
\]
Thus $C_0=A_0B_0+A_1B_3+A_2B_6$, etc.  Multiplication order is always left input form times right input form, so the construction is meaningful over noncommutative coefficient rings.

A rank-$R$ bilinear algorithm is a decomposition
\[
  \MM=\sum_{r=0}^{R-1} U_r\otimes V_r\otimes W_r,
\]
where each $U_r,V_r,W_r$ is a length-$9$ coefficient vector.  It computes
\[
  M_r=\Bigl(\sum_{i=0}^{8}U_{r,i}A_i\Bigr)\Bigl(\sum_{j=0}^{8}V_{r,j}B_j\Bigr),
  \qquad
  C_k=\sum_{r=0}^{R-1}W_{r,k}M_r .
\]
The displayed scheme has $R=23$ and all entries of $U,V,W$ are in $\{-1,0,1\}$.

The addition count is measured in the usual linear straight-line program model.  Inputs are free.  A gate computing $x+y$ or $x-y$ from previously available quantities costs one.  Copies and sign changes cost zero.  If a final output is assembled as a sum of $t$ available atoms, it costs $t-1$ additions.  Products $M_r$ are counted separately from additions.

Correctness of the factorization is equivalent to the Brent equations \cite{Brent1970}
\begin{equation}
\label{eq:brent}
  \sum_{r=0}^{22} U_{r,(i,k)}V_{r,(k',j)}W_{r,(i',j')}
  =\one[i=i']\one[j=j']\one[k=k']
\end{equation}
for all $i,j,k,i',j',k'\in\{0,1,2\}$.  There are $3^6=729$ equations.

\begin{table}[t]
\centering
\small
\begin{tabular}{@{}p{0.28\linewidth}ccp{0.16\linewidth}p{0.30\linewidth}@{}}
\toprule
Source & Rank & Additions & Coefficients & Presentation / verification \\
\midrule
Laderman~\cite{Laderman1976} & $23$ & $98$ & integer & printed noncommutative construction \\
Smirnov~\cite{Smirnov2013} & $23$ & $84$ naive & integer & tensor construction; additive schedule not optimized there \\
Stapleton~\cite{Stapleton2025} & $23$ & $60$ & ternary & printed scheme and Python verification script \\
M{\aa}rtensson--Wagner--Stapleton~\cite{MartenssonWagnerStapleton2025} & $23$ & $59$ & ternary & optimized scheme and factor-file appendix \\
Perminov~\cite{Perminov2025} & $23$ & $58$ & ternary & printed scheme and computational validation \\
This paper & $23$ & $56$ & ternary & printed scheme, integer Brent certificate, and public verifier \\
\bottomrule
\end{tabular}
\caption{Additive-complexity context for rank-$23$ algorithms for general $3\times3$ matrix multiplication.  ``Ternary'' means coefficients in $\{-1,0,1\}$.  Addition counts include additions and subtractions; older counts are included as summarized in recent comparison papers \cite{MartenssonWagnerStapleton2025,Perminov2025}.}
\label{tab:comparison}
\end{table}

\section{Discovery Method}
\label{sec:derivation}
We obtained the displayed scheme from a verified ternary rank-$23$ factorization of additive cost 58 due to Perminov~\cite{Perminov2025}. We applied the cyclic tensor automorphism

\[
   (U,V,W) \longmapsto (V,\,W^{T},\,U^{T}),
\]

where transposition is taken with respect to the $3\times 3$  row-major indexing. This automorphism preserves the matrix-multiplication tensor while permuting the roles of the left input, right input, and output recombination factors.

After this reorientation, correctness of the bilinear decomposition is inherited from the source factorization. The remaining task is to realize the transformed linear maps with fewer additions. The transformed left factor admits a $13$-addition vector-chain schedule inherited from the original construction. We then found a $13$-addition schedule for the transformed right factor and a $30$-addition output recombination schedule, consisting of $7$ shared product combinations and $23$ final-assembly additions. Thus the total additive cost is $13+13+30=56$.

\section{The 56-addition algorithm}
\label{sec:algorithm}

The straight-line program has three stages: first form input linear forms, then multiply corresponding left and right forms, and finally assemble the output entries.  All formulas below are in implementation order.  The input stage forms the following $13$ left-side intermediates and $13$ right-side intermediates:
{\small
\[
\begin{array}{rcl@{\qquad}rcl}
u_1&=&A_6-A_7 & v_1&=&B_2+B_5\\
u_2&=&A_4-u_1 & v_2&=&B_3-B_5\\
u_3&=&A_1-u_2 & v_3&=&B_1+B_4\\
u_4&=&u_3-A_6 & v_4&=&B_8-v_1\\
u_5&=&u_4-A_8 & v_5&=&B_0-v_3\\
u_6&=&A_1-u_5 & v_6&=&B_2-v_5\\
u_7&=&A_0-u_3 & v_7&=&B_4-v_6\\
u_8&=&u_5+A_2 & v_8&=&B_7-v_7\\
u_9&=&u_7-A_3 & v_9&=&B_2-v_8\\
u_{10}&=&u_9+A_5 & v_{10}&=&B_8+v_8\\
u_{11}&=&u_{10}-A_2 & v_{11}&=&v_2-v_6\\
u_{12}&=&A_0-A_1 & v_{12}&=&B_8+v_{11}\\
u_{13}&=&u_{12}-u_{10} & v_{13}&=&B_6-v_{12}.
\end{array}
\]
}

The product stage uses these intermediates together with raw input entries.  A leading minus sign denotes a zero-cost sign change.  Compute the $23$ bilinear products:
{\small
\[
\begin{array}{rcl@{\qquad}rcl}
M_0&=&u_6(-v_{11}) & M_{12}&=&u_7B_2\\
M_1&=&u_{13}(-v_7) & M_{13}&=&A_2B_6\\
M_2&=&A_5v_{10} & M_{14}&=&A_1B_3\\
M_3&=&u_8B_8 & M_{15}&=&A_6v_3\\
M_4&=&u_5(-v_{12}) & M_{16}&=&u_9v_9\\
M_5&=&A_4B_5 & M_{17}&=&A_0B_0\\
M_6&=&A_8B_7 & M_{18}&=&A_5B_6\\
M_7&=&(-u_2)v_6 & M_{19}&=&A_3B_0\\
M_8&=&A_4B_3 & M_{20}&=&(-u_{11})B_7\\
M_9&=&u_3v_5 & M_{21}&=&u_1B_4\\
M_{10}&=&(-u_4)(-v_4) & M_{22}&=&u_{10}(-v_8)\\
M_{11}&=&A_8v_{13}. &&
\end{array}
\]
}

The output stage first extracts seven shared combinations of products:
{\small
\[
\begin{array}{rcl@{\qquad}rcl}
w_1&=&M_4+M_9 & w_5&=&M_1-M_7\\
w_2&=&M_{12}+M_{22} & w_6&=&M_{10}+w_3\\
w_3&=&M_{14}+w_1 & w_7&=&M_{16}-w_2\\
w_4&=&M_0+M_7.&&
\end{array}
\]
}

Then assemble the nine output entries:
{\small
\[
\begin{array}{rcl@{\qquad}rcl}
C_0&=&M_{13}+M_{14}+M_{17} & C_5&=&M_2+M_5-w_7\\
C_1&=&-M_9+M_{17}+M_{20}-w_2+w_5 & C_6&=&-M_8+M_{11}+M_{15}+w_6\\
C_2&=&M_3+M_{12}+w_3+w_4 & C_7&=&M_6+M_{15}-M_{21}\\
C_3&=&M_8+M_{18}+M_{19} & C_8&=&-M_5+w_4+w_6\\
C_4&=&M_{19}+M_{21}+w_5+w_7.&&
\end{array}
\]
}

The final assembly costs are respectively $2,4,3,2,3,2,3,2,2$, summing to $23$.  Together with the $7$ shared output intermediates, the output side costs $30$ additions.

\section{Expanded bilinear presentation}
\label{app:expanded}

The following is the same rank-$23$ decomposition written without the shared input and output temporaries.  It is included for comparison with the compact algorithm tables in recent $58$- and $59$-addition papers.  This expanded presentation is not the $56$-addition implementation; the implementation order is Section~\ref{sec:algorithm}.  This appendix uses one-based matrix entries: $a_{ij}=A_{3(i-1)+(j-1)}$ and $b_{ij}=B_{3(i-1)+(j-1)}$.  The expanded products satisfy $p_{r+1}=M_r$ for $r=0,\ldots,22$.

\begin{Verbatim}[fontsize=\scriptsize]
p01 = (a22 + a32 + a33) * (-b11 + b12 + b13 - b21 + b22 + b23)
p02 = (a21 - a22 - a23 + a31 - a32) * (-b11 + b12 + b13)
p03 = (a23) * (-b11 + b12 + b13 + b32 + b33)
p04 = (a12 + a13 - a22 - a32 - a33) * (b33)
p05 = (a12 - a22 - a32 - a33) * (-b11 + b12 + b13 - b21 + b22 + b23 - b33)
p06 = (a22) * (b23)
p07 = (a33) * (b32)
p08 = (-a22 + a31 - a32) * (-b11 + b12 + b13 + b22)
p09 = (a22) * (b21)
p10 = (a12 - a22 + a31 - a32) * (b11 - b12 - b22)
p11 = (-a12 + a22 + a32) * (b13 + b23 - b33)
p12 = (a33) * (-b11 + b12 + b13 - b21 + b22 + b23 + b31 - b33)
p13 = (a11 - a12 + a22 - a31 + a32) * (b13)
p14 = (a13) * (b31)
p15 = (a12) * (b21)
p16 = (a31) * (b12 + b22)
p17 = (a11 - a12 - a21 + a22 - a31 + a32) * (b11 - b12 - b32)
p18 = (a11) * (b11)
p19 = (a23) * (b31)
p20 = (a21) * (b11)
p21 = (-a11 + a12 + a13 + a21 - a22 - a23 + a31 - a32) * (b32)
p22 = (a31 - a32) * (b22)
p23 = (a11 - a12 - a21 + a22 + a23 - a31 + a32) * (b11 - b12 - b13 - b32)

c11 = p14 + p15 + p18
c12 = p02 - p08 - p10 - p13 + p18 + p21 - p23
c13 = p01 + p04 + p05 + p08 + p10 + p13 + p15
c21 = p09 + p19 + p20
c22 = p02 - p08 - p13 + p17 + p20 + p22 - p23
c23 = p03 + p06 + p13 - p17 + p23
c31 = p05 - p09 + p10 + p11 + p12 + p15 + p16
c32 = p07 + p16 - p22
c33 = p01 + p05 - p06 + p08 + p10 + p11 + p15
\end{Verbatim}

\section{Compact ternary factor-file format}
\label{app:fileformat}

The following gives the tensor factors in a compact file format. The three blocks are $U^T$, $V^T$, and $W^T$, separated by a line containing \texttt{\#}.  In each block there are nine rows, one for the row-major coordinate $0,\ldots,8$, and each row contains the $23$ coefficients for products $M_0,\ldots,M_{22}$.  Thus the first block gives the coefficients of $A_0,\ldots,A_8$ in the left factors, the second block gives the coefficients of $B_0,\ldots,B_8$ in the right factors, and the third block gives the coefficients of $C_0,\ldots,C_8$ in the output recombination.

\begin{Verbatim}[fontsize=\scriptsize]
 0  0  0  0  0  0  0  0  0  0  0  0  1  0  0  0  1  1  0  0 -1  0  1
 0  0  0  1  1  0  0  0  0  1 -1  0 -1  0  1  0 -1  0  0  0  1  0 -1
 0  0  0  1  0  0  0  0  0  0  0  0  0  1  0  0  0  0  0  0  1  0  0
 0  1  0  0  0  0  0  0  0  0  0  0  0  0  0  0 -1  0  0  1  1  0 -1
 1 -1  0 -1 -1  1  0 -1  1 -1  1  0  1  0  0  0  1  0  0  0 -1  0  1
 0 -1  1  0  0  0  0  0  0  0  0  0  0  0  0  0  0  0  1  0 -1  0  1
 0  1  0  0  0  0  0  1  0  1  0  0 -1  0  0  1 -1  0  0  0  1  1 -1
 1 -1  0 -1 -1  0  0 -1  0 -1  1  0  1  0  0  0  1  0  0  0 -1 -1  1
 1  0  0 -1 -1  0  1  0  0  0  0  1  0  0  0  0  0  0  0  0  0  0  0
#
-1 -1 -1  0 -1  0  0 -1  0  1  0 -1  0  0  0  0  1  1  0  1  0  0  1
 1  1  1  0  1  0  0  1  0 -1  0  1  0  0  0  1 -1  0  0  0  0  0 -1
 1  1  1  0  1  0  0  1  0  0  1  1  1  0  0  0  0  0  0  0  0  0 -1
-1  0  0  0 -1  0  0  0  1  0  0 -1  0  0  1  0  0  0  0  0  0  0  0
 1  0  0  0  1  0  0  1  0 -1  0  1  0  0  0  1  0  0  0  0  0  1  0
 1  0  0  0  1  1  0  0  0  0  1  1  0  0  0  0  0  0  0  0  0  0  0
 0  0  0  0  0  0  0  0  0  0  0  1  0  1  0  0  0  0  1  0  0  0  0
 0  0  1  0  0  0  1  0  0  0  0  0  0  0  0  0 -1  0  0  0  1  0 -1
 0  0  1  1 -1  0  0  0  0  0 -1 -1  0  0  0  0  0  0  0  0  0  0  0
#
 0  0  0  0  0  0  0  0  0  0  0  0  0  1  1  0  0  1  0  0  0  0  0
 0  1  0  0  0  0  0 -1  0 -1  0  0 -1  0  0  0  0  1  0  0  1  0 -1
 1  0  0  1  1  0  0  1  0  1  0  0  1  0  1  0  0  0  0  0  0  0  0
 0  0  0  0  0  0  0  0  1  0  0  0  0  0  0  0  0  0  1  1  0  0  0
 0  1  0  0  0  0  0 -1  0  0  0  0 -1  0  0  0  1  0  0  1  0  1 -1
 0  0  1  0  0  1  0  0  0  0  0  0  1  0  0  0 -1  0  0  0  0  0  1
 0  0  0  0  1  0  0  0 -1  1  1  1  0  0  1  1  0  0  0  0  0  0  0
 0  0  0  0  0  0  1  0  0  0  0  0  0  0  0  1  0  0  0  0  0 -1  0
 1  0  0  0  1 -1  0  1  0  1  1  0  0  0  1  0  0  0  0  0  0  0  0
\end{Verbatim}

\section{Reproducibility}
\label{app:reproducibility}

The following public GitHub repository contains our reference implementation of this algorithm and the verification code described in this paper: \url{https://github.com/sunyinqi0508/3by3r23-56a}

The verifier reconstructs the tensor factors $U,V,W$ from the printed straight-line program and checks that they match the compact ternary factor format in Appendix~\ref{app:fileformat}. It then recomputes the straight-line costs $13$, $13$, and $30$, verifies all $729$ Brent equations over $\mathbb{Z}$, and repeats the checks modulo $2$ and $3$. The finite-field checks are redundant after the integer check, but they are useful for catching implementation errors such as sign mistakes or coordinate transpositions. The verifier also runs randomized end-to-end tests over bounded integer inputs, over $\mathbb{F}_2$ and $\mathbb{F}_3$, and over a noncommutative test ring represented by random $2\times2$ integer matrices.

\end{document}